# Antineutrino Spectrum of the Earth and the Problem of Oscillating Geoantineutrino Deficit

V.D. Rusov[1*], V.N. Pavlovich[2], V.N. Vaschenko[3-4], V.A. Tarasov[1], D.A. Litvinov[1],
V.N. Bolshakov[1], E.N. Khotyaintseva[2]

[1]*Odessa National Polytechnic University, 65044 Odessa, Ukraine*
[2]*Institute for Nuclear Researches of National Academy of Science of Ukraine, 01028 Kiev, Ukraine*
[3]*Ukrainian Antarctic Center, 01033 Kiev, Ukraine*
[4]*National University by T. Shevchenko, 01017 Kiev, Ukraine*

(December 19, 2003)

Based on O'Nionse-Evenson-Hamilton model of geochemical evolution of the mantle differentiation and the earth's crust growth the integral estimations of geoantineutrino flow intensity $\Phi_{\tilde{\nu}}(^{238}U) \approx 2.2 \cdot 10^6$, $\Phi_{\tilde{\nu}}(^{232}Th) \approx 1.8 \cdot 10^6$ and $\Phi_{\tilde{\nu}}(^{40}K) \approx 9.12 \cdot 10^6$ $cm^{-2}s^{-1}$ on the Earth's surface in the absence of oscillations are obtained. According to the obtained estimations the partial and total $d\Phi_{\tilde{\nu}}/dE$ ($^{238}U+^{232}Th+^{40}K$) antineutrino energy spectra are constructed. Using the partial $d\Phi_{\tilde{\nu}}/dE$ ($^{238}U$), $d\Phi_{\tilde{\nu}}/dE$ ($^{232}Th$) and total $d\Phi_{\tilde{\nu}}/dE$ ($^{238}U+^{232}Th$) energy spectra in absence of oscillations and taking into account the inverse $\beta$-decay reaction threshold ($E_{\tilde{\nu}}^{thr} = 1.804$ MeV) and KamLAND detector characteristics partial and total energy spectra of detected by KamLAND detector antineutrinos are obtained. It is shown that the integral of these spectra with allowance for oscillations (under condition of a transition probability ($P_{\tilde{\nu} \to \tilde{\nu}} \cong 0.85$) is in 3 times less than a similar integral of best fit of experimental reactor antineutrino KamLAND-spectrum obtained on the assumption of 9 geoantineutrinos detection.

The reasons of occurrence of the problem of oscillating geoantineutrino deficit and latent radiogenic heat-evolution power of the Earth accordingly are considered.

## 1. Introduction

Now it is obvious that experiments on antineutrino sounding of the Earth interior [3] by the multidetector scheme of geoantineutrino detection on large base [4] can give forcible arguments in favour of existing theoretical notions of spatial-temporal radiogenic energy sources distribution in models of the Earth geochemical evolution [1] and reactor antineutrino oscillation in LMA-area [2]. Obtained in such a way geoantineutrino spectra make it possible to solve direct problem of geoantineutrino flow determination on Earth surface and "to grope" the solution of inverse problem of $\beta$-sources radial profile restoration in the Earth's interior with allowance for the correlation function and statistics of geoantineutrino detection [5].

Adduced in Refs. [6, 7] the trial estimations of geoantineutrino flow corresponding to different models of the Earth internal composition, in our opinion, are not reliable as ignore the impotent features of temporal change of total radiogenic heat evolution of the Earth [1]. In particular, the present-day experimental measured geothermal flow power $H \approx 40$ TW in reality is characterized not current geoantineutrino flow intensity, but its value in distant past corresponding to thermal relaxation time of the Earth, which by the estimations of Ref. [1] amounts approximately $2 \cdot 10^9$ years. On the other hand, used in Refs. [7-9] simplified models of an internal composition of the Earth (chondrite [7] and silicate [7-9]) leave out also such important processes of geochemical evolution of the Earth as mantle differentiation and the crust growth, which reflect developmental transport processes of elements from so-called depleted mantle to the earth's crust area. The neglect of these problems renders impossible the determination the modern deep distribution of radioactive elements $^{238}$U, $^{232}$Th, $^{40}$K in the mantle and the earth's crust and corresponding latent integral heat-evolutions, which, in essence, qualitatively and quantitatively determine the reliability of partial and total estimations of flow intensity of the Earth's antineutrino radiation.

---

[*] Corresponding author: E-mail: siiis@te.net.ua

Obtaining of energy spectra and an integral estimation of geoantineutrino intensity on Earth's surface from various radioactive sources ($^{238}$U, $^{232}$Th and $^{40}$K) by the analysis of temporal evolution of radiogenic heat-evolution power of the Earth within the framework of model of geochemical processes of mantle differentiation and the earth's crust growth [1] is the purpose of the present paper.

## 2. Analysis of O'Nionse-Evenson-Hamilton model of geochemical evolution

Used in the present paper the reference results were obtained by O'Nionse, Evenson and Hamilton in 1979 as a result of geochemical simulation of mantle differentiation and the earth's crust growth [1]. Let us briefly consider these results and estimate the latent radiogenic heat flux.

The evolutionary geochemical model construction was based on data for daughter products $^{40}$Ar, $^{87}$Sr, $^{143}$Nd, $^{208}$Pb, $^{207}$Pb and $^{206}$Pb of radioisotopes $^{40}$K, $^{87}$Rb, $^{147}$Sm, $^{232}$Th, $^{235}$U and $^{238}$U accordingly. Now there are numerous experimental data concerning these isotopes in earth's crust and atmosphere distribution. The meteorites and basalts data make it possible to judge about primary composition of the Earth and mantle, from which the basalts are smelted. The model is under construction so that the observed ratio of isotopes $^{40}$Ar/$^{36}$Ar, $^{87}$Sr/$^{86}$Sr, $^{143}$Nd/$^{144}$Nd, $^{208}$Pb/$^{204}$Pb, $^{207}$Pb/$^{204}$Pb and $^{206}$Pb/$^{204}$Pb in earth's crust, atmosphere and mantle was satisfied.

It is supposed that the Earth was formed $4.55 \cdot 10^9$ years ago and had chemically homogeneous silicate component. From this moment an external 50-km layer of the Earth (reservoir $L$) began to form from the primary silicate reservoir (the mantle $M$) due to gradual entry of radioactive nuclides and elements with close physico-chemical properties, which are dissolved in rocks of earth's crust better than in ultra-based silicates of mantle, from M. It is considered that the continental crust formation happened during, at least, last $3.8 \cdot 10^9$ years (age of most ancient rocks).

The evolutionary model based on the balance equation of rate of each $i$-th component change in reservoirs L and M. So, if $n_i^L$ is mol number of $i$-components in the reservoir $L$, the rate of change $n_i^L$ is described by first-order differential equation:

$$\frac{dn_i^L}{dt} = \alpha_{ML}(t) \cdot n_i^M - \beta_{LM}(t) \cdot n_i^L + \lambda_j \cdot n_j^L, \qquad (1)$$

where $n_i^M$ is mol number of $i$-th component in the reservoir $M$; $\alpha_{ML}(t)$ and $\beta_{LM}(t)$ are transport coefficients of $i$-th component from $M$ to $L$ accordingly; $\lambda_j$ is radioactive decay constant of $n_j^L \to n_i^L$ (if it exists). Note that the transport coefficients are exponential functions of time (just as radiogenic heat release in the interior of the Earth, which causes a convection in the mantle):

$$\alpha_{ML}(t) = \alpha_0 \exp(-t/\tau_\alpha), \quad \beta_{LM}(t) = \beta_0 \exp(-t/\tau_\beta), \tag{2}$$

where $\tau_\alpha \sim 2 \cdot 10^9$ and $\tau_\beta \sim 0.65 \cdot 10^9$ years are characteristic time of *i*-component transport from $M$ in $L$ and vice versa respectively; preexponential factors $\alpha_0$ and $\beta_0$ have unlike signs for each element, but are identical for isotopes of the same element; factor $\alpha_{0i}$ is selected so that to obtain known concentration of *i-th* component in $L$ at $t=4.55 \cdot 10^9$ years; $\beta_{0i}$ is identical for all elements excepting Ar, which percolates in atmosphere.

Building the evolutionary model O'Nionse, Evenson and Hamilton have set the abundances of dispersed elements K, Ar, Rd, Sr, Sm, Nd, U, Th, Pb and isotopic composition of Ar, Sr, Nd and Pb in undifferentiated mantle, which were $4.55 \cdot 10^9$ years ago (Table I). For the description of the Earth's present elemental composition, under which the theory are adjusted, the known at present data concerning the abundances of elements in an external 50-km layer and in the residual (depleted) mantle, from which the part of elements has transformed in the reservoir $L$ (Table II), were collected.

Simulating the system of differential equations (1)-(2) solutions, describing the geochemical evolution from the Earth's primary composition (Table I) to a present composition of the earth's crust and all reservoir $L$ (Table II), O'Nionse, Evenson and Hamilton have shown that the optimum accordance with experimental data is achieved, when third of mantle mass appears in $M$ only. Zharkov [1], in its turn, has shown that the best accordance of such calculations with the observation data is achieved, if the reservoir $M$ contains only ~ 1/4 mantles mass, i.e. only upper mantle (~ 700 km). In any case the geochemical simulation has shown that considerable exchange between upper and bottom mantle during history of the Earth, apparently, was not. It means that the earth's crust was formed as a result of thermal and chemical evolution of the upper mantle, consequently the residual upper mantle is depleted of dispersed elements, and the bottom mantle haves the primary composition (Table I). Geometry and distribution parameters of heat flux in the mantle $M$, depleted mantle $LM$ and the crust $L$ corresponding to model calculations of O'Nionse, Evenson and Hamilton (with Zharkov's correction [1]) are shown in Fig. 1.

Using the Table I data it is possible to calculate integral heat-evolution in the silicate reservoir of the Earth ($M+L$) (see Fig. 2). As shown in Fig. 2 latent radiogenic heat-evolution power of the Earth is ~19.5 TW (Fig. 1, curve U+Th+K). Hence, if to suppose that the loss of heat is the result of radiogenic heat, the thermal relaxation time of the Earth approximate $2.5 \cdot 10^9$ years (in Fig. 1 it is shown by right arrow on time axis). Taking into account that the heat flow from a core in mantle (unrelated with a radioactive decay) is ~ 1/7 of total heat flow of the Earth [1], it is possible to consider that thermal relaxation time of the Earth decreases approximately up to $\sim 2 \cdot 10^9$ years. Thus due to large thermal inertia of the Earth, i.e. low heat conductivity, heat generated in the Earth's interior is transferred to a surface not instantly, but with delay for thermal relaxation time.

In reality the heat flow from planet interior can have component caused directly by primary heat, which has arisen at planet formation and as a result of following gravitational differentiation into core and mantle. Now this part of heat flow cannot be estimated [1].

## 3. Estimation of geoantineutrino integral intensity (no oscillation)

Let us determine geoantineutrino flow intensity on the Earth's surface produced by radioactive sources $^{238}$U, $^{232}$Th and $^{40}$K:

$$^{238}U \rightarrow {}^{206}Pb + 8\,{}^{4}He + 6e + 6\tilde{\nu}, \tag{3}$$

$$^{232}Th \rightarrow {}^{208}Pb + 6\,{}^{4}He + 4e + 4\tilde{\nu}, \tag{4}$$

$$^{40}K \rightarrow {}^{40}Ca + e + \tilde{\nu}. \tag{5}$$

It is easy to show that on the surface of the Earth ($R_\oplus \sim 6400$ km) non-oscillating geoneutrino intensity radiated by intra-earth radioactive sources depends on geometry like [7]:

$$\Phi_{\tilde{\nu}} = \frac{G}{4\pi R_\oplus^2} L_{\tilde{\nu}}, \tag{6}$$

where the geometrical factor $G$ for spherical homogeneous shells with radiuses $r_1=x_1 R_\oplus$ and $r_2=x_2 R_\oplus$ is equal [7]:

$$G = \frac{3}{2(x_2^3 - x_1^3)} \left\{ \frac{1}{2}(1-x^2)\ln\frac{1+x}{1-x} - x \right\}\Bigg|_{x_2}^{x_1}, \tag{7}$$

and $L_{\tilde{\nu}}$ is additive intensity of geoantineutrino (in units of $10^{24}$ s$^{-1}$) radiated according Eqs. (3)-(5) by natural uranium, thorium and potassium (in units of $10^{17}$ kg) [7]:

$$L_{\tilde{\nu}} = 7.41 M(U)\left[1 + 0.22\,Th/U + 0.36\cdot 10^{-3}\,K/U\right], \tag{8}$$

Obviously, to determine geoantineutrino intensity on the surface of the Earth (8) it is necessary to know the natural uranium mass $M_i(U)$ and value of ratio Th/U and K/U in different subsystems, i.e. geospheres (spherical shells of the Earth), on which the model system is broken. In considered model [1] the depth profile has three geospheres, i.e. crust, depleted mantle (from which the earth's crust was formed) and primary (undifferentiated) mantle. In this case for definition $M_i(U)$ (in units of $10^{17}$ kg) the expression for a geothermal flow $H$ formed by a homogeneous spherical shell [7] was used:

$$H = 9.5 \cdot M(U)\left[1 + 0.284 \cdot Th/U + 3.8\cdot 10^{-5}\,K/U\right], \quad [TW], \tag{9}$$

where for the mantle $(Th/U)_{mantle} = 4$ and $(K/U)_{mantle} = 10^4$ are determined by parameters of a primary composition (see Table I), for the earth's crust $(Th/U)_{crust} = 2.5$ and $(K/U)_{crust} = 1.23\cdot 10^4$ are determined by parameters of present composition (see Table II), whereas for the depleted mantle $(Th/U)_{depl} = 9.9$ and $(K/U)_{depl} = 0.157\cdot 10^4$ are predetermined by a peculiar conservation law of primary composition, i.e. sum of masses of depleted mantle and earth's crust:

$$\frac{(m(*) \cdot V \cdot \rho)_{depl} + (m(*) \cdot V \cdot \rho)_{crust}}{(m(U) \cdot \rho)_{mantle} \cdot V_{depl}} = \begin{cases} 4 & для \quad m(*) \equiv m(Th), \\ 10^4 & для \quad m(*) \equiv m(K), \end{cases} \quad (10)$$

$$(m(U) \cdot \rho)_{mantle} \cdot V_{depl} = M_{depl}(U) + M_{crust}(U), \quad (11)$$

where $M_{depl}(U)$, $M_{crust}(U)$ and $m(U)$, $m(Th)$ and $m(K)$ are masses (kg) and abundance (g/ton=$10^{-6}$ kg/kg) of elements dispersed in appropriate geospheres; $V$ and $\rho$ are volume and density of medium in appropriate geospheres: the mantle (index "mantle"), depleted mantle (index "depl") and earth's crust (index "crust").

We have obtain the geometrical factor $G_i$ and masses of natural uranium, thorium and potassium by Eqs. (7) and (9), and also estimation of geoantineutrino intensity on Earth's surface by Eq. (6) (see Table III). The values of parameters, for example, bedding geometry, density $\rho$ and specific heat-evolution of geosphera medium $q$ are borrowed from O'Nionse, Evenson and Hamilton model and are indicated in Fig. 2.

Let us compare the obtained theoretical estimations of antineutrino integral intensity on the Earth's surface emitted by natural uranium ($\Phi_{\tilde{\nu}}(^{238}U) \approx 2.2 \cdot 10^6 \ cm^{-2}s^{-1}$) and thorium ($\Phi_{\tilde{\nu}}(^{232}Th) \approx 1.8 \cdot 10^6 \ cm^{-2}s^{-1}$) with experimental data of KamLAND Collaboration [11].

## 4. The geoantineutrino energy spectra (no oscillations) and KamLAND experiment

The knowledge of antineutrino total energy spectrum emitted by fission products in reactor core and background caused by geoantineutrino total energy spectrum of radioisotopes $^{238}$U, $^{232}$Th, $^{40}$K is necessary for interpretation of neutrino experiments at nuclear reactors [2, 11].

The most irrefragable and physically grounded characteristic of calculated $\tilde{\nu}$-spectrum of $\beta$-radioactive nuclide are the total emission $\beta$-spectrum obtained by partial $\beta$-spectra weighing on intensity of allowed and forbidden $\beta$-transitions [12].

Total $\tilde{\nu}$-spectrum results from total $\beta$-spectrum of some nuclide by calculation of partial $\tilde{\nu}$-spectra on the basis of previously calculated partial $\beta$-spectra reasoning from electrons and antineutrino kinematical connection, i.e. $E_{\tilde{\nu}} = E_0 - E_\beta$ (where $E_0$ is total energy distributed between an electron ($E_\beta$) and antineutrino ($E_{\tilde{\nu}}$)). In other words, the energy conservation law predetermines a way of $\tilde{\nu}$-spectrum finding of any partial transition. It is dissymmetrical to the partial $\beta$-spectrum, i.e. $N_{\tilde{\nu}}(E_{\tilde{\nu}}) = N_\beta(E_\beta - E_{\tilde{\nu}})$. It makes possible to obtain $\tilde{\nu}$-spectrum by "mirror overturn" and, that is most important, with accuracy like calculation and/or measured $\beta$-spectrum. Therefore the forecast precision of effective total spectrum of radioactive actinoid is equally determined by the nuclear data accuracy in the used catalog of $\beta$-decay characteristics, partial $\beta$- and $\tilde{\nu}$-spectra and data correctness of the same radioactive decay products storage [12].

Using the calculation procedures of partial and total energy $\beta$-, $\tilde{\nu}$-spectra [12] and ENSDF library [13] we have constructed the partial d $d\Phi_{\tilde{\nu}}/dE$ ($^{238}$U), $d\Phi_{\tilde{\nu}}/dE$ ($^{232}$Th), $d\Phi_{\tilde{\nu}}/dE$ ($^{40}$K) (Fig. 3a-c) and total energy spectra of the Earth's $d\Phi_{\tilde{\nu}}/dE$ ($^{238}$U+$^{232}$Th+$^{40}$K) (Fig. 3d) (the partial contributions were previously normalized on corresponding geoantineutrino integral intensity on the Earth's surface (see. Table III)).

Then using obtained partial $d\Phi_{\tilde{\nu}}/dE$ ($^{238}U$), $d\Phi_{\tilde{\nu}}/dE$ ($^{232}Th$) and total $d\Phi_{\tilde{\nu}}/dE$ ($^{238}U+^{232}Th$) energy spectra and taking into account the detection threshold ($E_{\tilde{\nu}}^{thr}$ =1.804 MeV) in inverse $\beta$-decay reaction and also KamLAND detector characteristics, we have obtain the theoretical form of measured total energy spectrum $dn_{\tilde{\nu}}/dE$ by basic equation of an antineutrino spectrometry [5] describing the contribution of each radioactive isotopes (at given geometry and detector characteristics) (Fig. 4):

$$\frac{dn_{\tilde{\nu}}}{dE} = \varepsilon \cdot N_p \cdot \sum_i \frac{d\lambda_{\tilde{\nu}}^i}{dE} \cdot \sigma_{vp}(E_{\tilde{\nu}}) \cdot \Delta t, \quad MeV^{-1}, \qquad (14)$$

where $d\lambda_{\tilde{\nu}}/dE \equiv d\Phi_{\tilde{\nu}}/dE$ at $E_{\tilde{\nu}} \geq 1.804$ MeV (Fig. 4); according to Ref. [11] $\varepsilon \approx 0.783$ is detector efficiency; $N_P=3.46 \cdot 10^{31}$ is proton number in the detector sensitive volume; $\Delta t= 1.25 \cdot 10^7$ s is exposure time; $\sigma_{vp}$ is antineutrino-proton interaction cross-section for the inverse $\beta$-decay reaction, which with the allowance for its behavior close to reaction threshold ($\delta_{thr}$), recoil ($\delta_{rec}$), weak magnetism ($\delta_{WM}$) and radiation corrections ($\delta_{rad}$) looks like [14, 15]:

$$\sigma_{vp}(E_{\tilde{\nu}}) = \sigma_0(E_{\tilde{\nu}}) \cdot (1+\delta_{thr}) \cdot (1+\delta_{WM} +\delta_{rec}) \cdot (1+\delta_{rad}), \qquad (15)$$

The analytical expressions for all corrections and their detailed discussion are given in Refs. [14, 15]. The "naive" cross-section $\sigma_0(E_{\tilde{\nu}})$ corresponding to infinitely heavy nucleons approximation or $E_n \approx m_n$, $E_{e+} \ll m_n$ looks like

$$\sigma_0(E_{\tilde{\nu}}) = \frac{2\pi^2 \hbar^3 \ln 2}{m_e^5 c^7 (f\tau_{1/2})} \cdot \frac{1}{c}\left[(E_{\tilde{\nu}} - \Delta)^2 - m_e^2 c^4\right]^{1/2} \cdot (E_{\tilde{\nu}} - \Delta), \qquad (16)$$

where $E_{\tilde{\nu}} - (m_n - m_p)c^2 = E_{\tilde{\nu}} - \Delta = E_{e+}$ is total positron energy in the inverse $\beta$-decay reaction; ($f \cdot \tau_{1/2}$) is so-called reduced neutron half-life [12], for which phase space factor of neutron $f$=1.7146 is determined to 0.01 % [16] and half-life is $\tau_{1/2}= \tau \cdot \ln 2$, (where $\tau$= 887.4±0.2% s) [17].

The quantitative estimation of KamLAND-experiment data by the integrals (14) of partial energy spectra $dn_{\tilde{\nu}}/dE$ ($^{238}U$) and $dn_{\tilde{\nu}}/dE$ ($^{232}Th$) (Fig. 5a, b) or $dn_{\tilde{\nu}}/dE$ ($^{238}U+^{232}Th$) (Fig. 5c):

$$n_{\tilde{\nu}} = \varepsilon \cdot N_p \cdot \Delta t \cdot \int_{E=1.804}^{\infty} \frac{d\lambda_{\tilde{\nu}}(U+Th)}{dE} \cdot \sigma_{vp}(E)dE \cong 2.70(U)+0,78(Th) \cong 3.5 \qquad (17)$$

shows that without taking oscillations it is in 2,6 times less and taking into account neutrino oscillations (under condition of transition probability $P_{\tilde{\nu} \to \tilde{\nu}} \cong 0.85$ [18]) it is approximately in 3 times less than similar integral of reactor antineutrino experimental KamLAND spectrum obtained on the assumption of 9 geoantineutrinos detection ($n_{\tilde{\nu}} \approx 4$ from $^{238}U$ and $n_{\tilde{\nu}} \approx 5$ from $^{232}Th$).

So, the comparative analysis of theoretical and experimental neutrino data, strange as it may seem, engenders in new form a known problem of shortages, but now not of solar neutrino and of oscillating geoantineutrino. And that is why.

If to consider the number of geoantineutrino ($\approx 9$) "detected " in KamLAND-experiment as a reference point for modern models of antineutrino spectrum of the Earth, the increase of the mass of natural uranium and thorium (i.e. power of the Earth's heat flow) in 2,6 times (in the absence of oscillations) or in 3 times (in the presence of oscillations) is impossible due to following physical limitations.

Since 60-years and till now about 10 thousands high-precision measurements of heat flow power $H$ is accumulated and their statistical accuracy makes it impossible to manipulate value $H$ within the framework of so-called geophysical uncertainty such as "the factor of 2" [7]. On the other hand, the maximally possible radiogenic heat power can not to exceed 22-23 TW as this value provides minimally possible value of thermal relaxation time of the Earth ($\sim$ 1 mlrd. years) (see Fig. 1). If to add, that only in extremal model [7, 9] (in which the total heat flow of the Earth ($\sim$ 40 TW) is identified with radiogenic flow) the calculated total antineutrino flow ($\Phi_{\tilde{\nu}}(^{238}U) \approx 5.72 \cdot 10^6$ $cm^{-2}s^{-1}$ and $\Phi_{\tilde{\nu}}(^{232}Th) \approx 5.03 \cdot 10^6$ $cm^{-2}s^{-1}$) produces total number of events hypothetically detected by the KamLAND-detector [18, 9] approximately equal to a reference point ($\approx 9$), i.e.

$$N_{osc} \cong (1 - 0.5\sin\theta_{12}) \cdot N_{no} \cdot \frac{\varepsilon \cdot N_p \Delta t}{10^{32} \tau_{year}} = 0.85 \cdot N_{no} \cdot \frac{0.783 \cdot 3.46 \cdot 10^{31} \cdot 1.25 \cdot 10^7}{10^{32} \cdot 3.156 \cdot 10^7} \approx 8.8,$$

$$N_{no} = 13.2 \cdot 10^{-6} \cdot \Phi_{\tilde{\nu}}(u) + 4.0 \cdot 10^{-6} \cdot \Phi_{\tilde{\nu}}(Th) \cong 96,$$

it means that the geoantineutrino deficit is stably observed in modern models.

### 5. Discussion and conclusions

The estimations of integral intensity and energy spectrum of geoantineutrino on the Earth's surface (in the absence of oscillations) from different radioactive sources ($^{238}U$, $^{232}Th$ and $^{40}K$) are obtained on the basis of temporal evolution analysis of radiogenic heat-evolution power of the Earth within the framework of model of geochemical processes of mantle differentiation and earth's crust growth [1].

Note that the convergence of obtained estimations of "detected" geoantineutrino ($n_{\tilde{\nu}}$ =3.5 from $^{238}U$ и $^{232}Th$) on the surface of the Earth from natural uranium ($\Phi_{\tilde{\nu}}(^{238}U) \approx 2.2 \cdot 10^6$ cm$^{-2}$s$^{-1}$) and thorium ($\Phi_{\tilde{\nu}}(^{232}Th) \approx 1.8 \cdot 10^6$ cm$^{-2}$s$^{-1}$) with similar estimations of KamLAND Collaboration determines a nontrivial problem of stable (concerning existing geomodels) deficit of oscillating geoantineutrino. At the same time this conclusion, in our opinion, should not mask other grave problems of description of partial and total antineutrino spectra of the Earth and KamLAND-experiment observation data interpretation.

It is explained in the following way. Obviously that disadvantages of models of the Earth's geochemical evolution (from a primary to modern composition), which determine error of geo(anti)neutrino integral intensity estimation, are sequel of while insuperable difficulties of most advanced (parametric, rheological and physical) models of an internal composition of the Earth [1]. The inverse path, i.e. searches of optimum models of the Earth's geochemical evolution by comparison or adjustment of antineutrino integral intensity obtained by these models with similar estimations of KamLAND-experiment observation data, is not well-founded due to evident weakness of estimation procedure of constant bias of KamLAND-experiment some parameters. First of all it concerns such interdependent magnitudes as nuclear fuel composition and antineutrino total spectrum. The uncertainty of their values is mainly caused by error of one parameter, i.e. the contribution from isotopes $\alpha_i$ to nuclear fuel composition.

As is well known, the contributions from isotopes $\alpha_i$ are usually calculated by the results of radiochemical analysis of spent nuclear fuel and in this case, for example, the error is approximately 5% for WWER-400 reactor (WPR type reactor) [19]. Moreover, because of $^{235}$U burnup and plutonium accumulation the contributions from isotopes $\alpha_i$ change noticeable during campaign, therefore it is difficult to believe that the error of fuel composition determination in KamLAND-experiment is $\approx$ 1 % [11], especially, if to take into account that the antineutrino energy spectrum changes can amount to 10% and the total cross-section of inverse $\beta$-decay reaction changes by almost 6% during one campaign (as it was experimentally shown [19]).

With the regard for these details concerning, apparently, not qualitative but quantitative results KamLAND experiment can lead to the refinement of oscillation parameters corresponding to best fit of the observation data.

The considered problem resolution, in our opinion, can become the results of independent experiment on geo(anti)neutrino energy spectrum measurement by the multidetector scheme of geo(anti)neutrino detection on large base [4]. At the same time the solutions of direct and inverse problem of neutrino remote diagnostic [5] of terrestrial processes for the measurement of pure geo(anti)neutrino spectrum and correct determination of radial profile of $\beta$-sources distribution in interior of the Earth, beyond doubt, will help partially or completely to solve a problem of geo(anti)neutrino deficit.

At last let us remind that the known breakthroughs in new physics area were more than once connected with neutrino [20]. In this sense the results of two-flavor analysis of KamLAND-experiment observation data have yielded not only a reactor antineutrino oscillation parameters compatible with values of corresponding LMA-solutions for a solar neutrino [11], but for the first time have paid our attention to the extremely important area of researches as experimental and theoretical geo(anti)neutrino physics, which is the part of massive mixing neutrino new physics and (that is specially important) "without inertia" component of such area of science as physics of the Earth.

**TABLES**

TABLE I. Abundances of dispersed elements in the Earth (in crust and mantle [$L+M$]).

| Element | Abundance* (g/ton) | Heat production rate (mcW/kg) |
|---|---|---|
| K | 200 | $3{,}6 \cdot 10^{-3}$ |
| Rb | 0,67 | - |
| Sr | 21,0 | - |
| U | 0,02 | 95,0 |
| Th | 0,08 | 27,0 |
| $Pb^0$ ** | 0,10 | - |
| Sm | 0,32 | - |
| Nd | 0,97 | - |
| Rb/Sr | 0,03 | - |
| K/U | $10^4$ | - |
| K/Rb | 300 | - |
| Sm/Nd | 0,33 | - |
| Th/U | 4 | - |
| Undifferentiated mantle | - | $4{,}8 \cdot 10^{-6}$ |

* The abundances are obtained on the assumption that the Earth has identical with carbonaceous chondrites relative abundances with respect to Ca, U, Th, Sm and Nd (O'Nionse, Evenson and Hamilton [1]).
** $Pb^0$ is primary lead.+

TABLE II. Abundances of dispersed elements (g/ton) in continental crust and outer 50-km layer (reservoir $L$).

| Element | Continental crust model [10] | Reservoir composition* |
|---|---|---|
| K | 12400 | 3800 |
| Pb | 50 | 15,3 |
| Sr | 400 | 122 |
| Sm | 3,7 | 1,13 |
| Nd | 16,0 | 4,90 |
| U | 1,0 | 0,31 |
| Th | 2,5 | 0,77 |
| $Pb^0$ ** | 5,0 | 1,5 |
| $^{87}Sr/^{86}Sr$ | - | 0,7120 |

* The estimation of 50-km layer modern composition without taking oceanic sediment layer, whose inclusion can increase abundance of $K$ no more than on 200 g/ton.
** $Pb^0$ is primary lead.

TABLE III. Mass distribution, antineutrino fluxes and heat production rates
($M$, $\Phi$ and $H$ are in units of $10^{17}$ kg, $10^6$ cm$^{-2}$s$^{-1}$ and TW respectively)

| Geo-spheres | $G$ | $^{238}$U | | | $^{232}$Th | | | $^{40}$K | | | $H$ |
| --- | --- | --- | --- | --- | --- | --- | --- | --- | --- | --- | --- |
| | | $^iM$ | $\Phi_{\tilde{\nu}}$ | $H$ | $^iM$ | $\Phi_{\tilde{\nu}}$ | $H$ | $^iM$ | $\Phi_{\tilde{\nu}}$ | $H$ | |
| Crust | 3.27 | 0.22 | 1.040 | 2.10 | 0.55 | 0.57 | 1.50 | 0.271 | 4.60 | 0.97 | 4.6 |
| Depleted mantle | 1.95 | 0.06 | 0.170 | 0.60 | 0.59 | 0.36 | 1.60 | 0.0094 | 0.95 | 0.04 | 2.2 |
| Mantle | 1.30 | 0.53 | 0.992 | 5.10 | 2.10 | 0.87 | 5.70 | 0.53 | 3.57 | 1.90 | 12.7 |
| Σ | - | 0.81 | 2.20 | 7.80 | 3.24 | 1.80 | 8.80 | 0.81 | 9.12 | 2.90 | 19.5 |

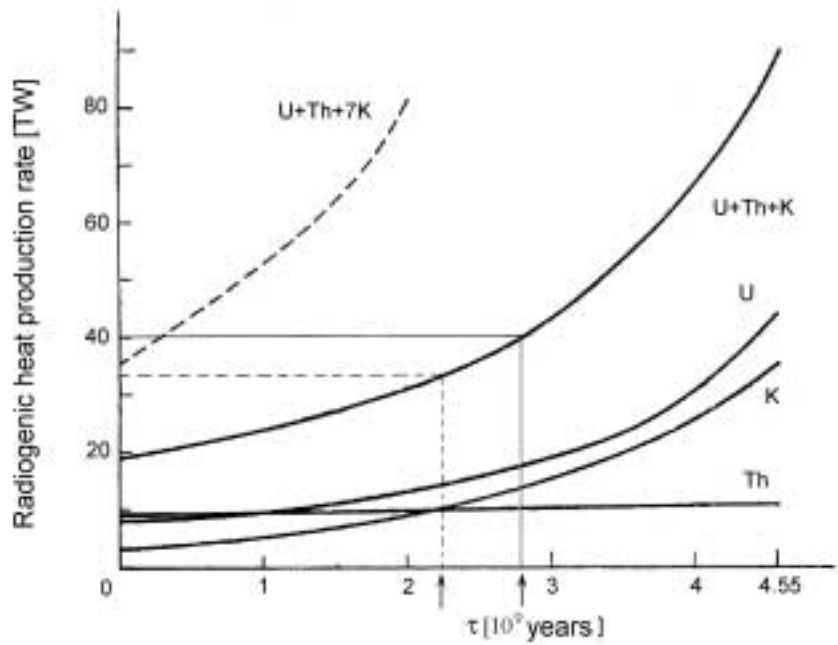

Fig. 1. Total radiogenic heat production rate of the Earth according to data of the Table II. Heat production caused by K, Th, U decay and their sum K+Th+U is separately shown. The curve U+Th+7K shows heat production for limiting chondrite concentration of potassium, i.e. K/U ≈ 7·10⁴. The value of modern loss of Earth's heat 40 TW (solid line) and loss of heat diminished on heat flow from a core to the mantle equal to 34,3 TW (hatch line) are shown also. The arrows on time axis indicate the dead time, i.e. thermal relaxation time of the Earth taking into account and without taking heat flow from a core to the mantle.

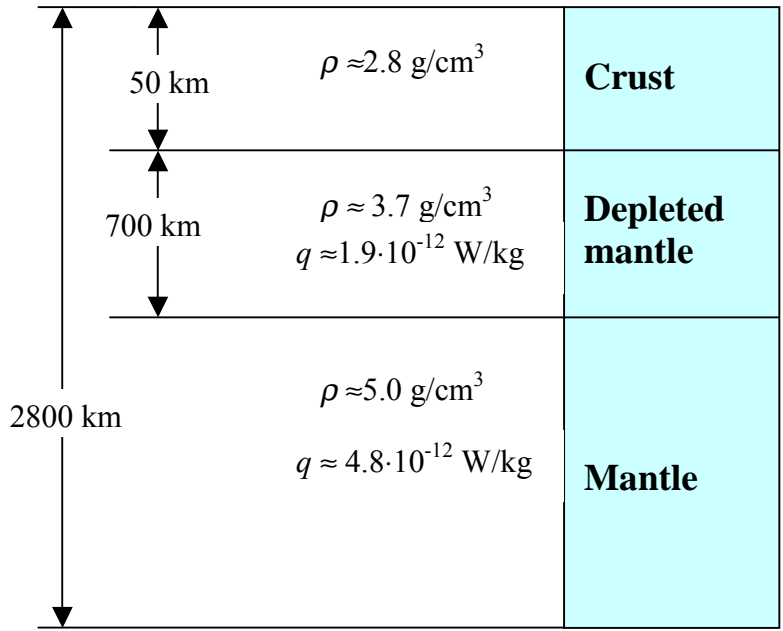

Fig. 2. The bedding geometry, density $\rho$ and specific heat-evolution of geosphera medium $q$ (O'Nionse-Evenson-Hamilton model).

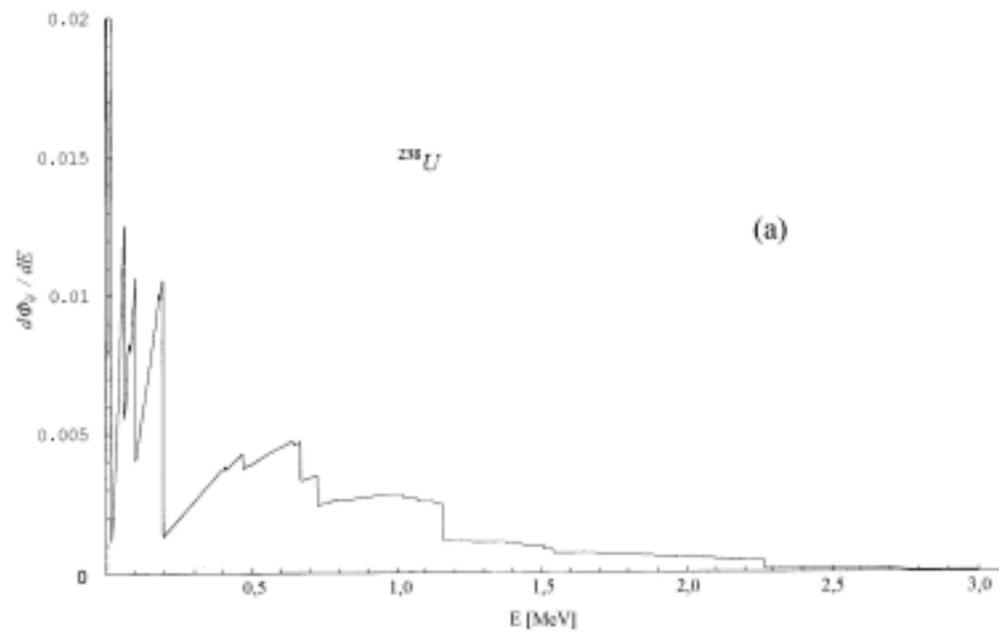

(a)

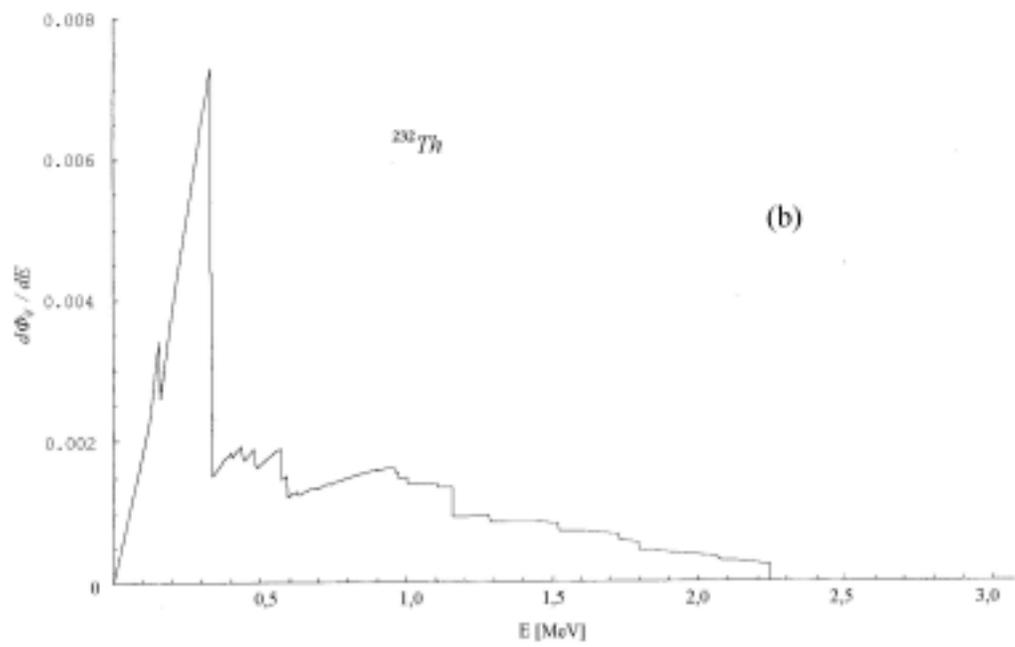

(b)

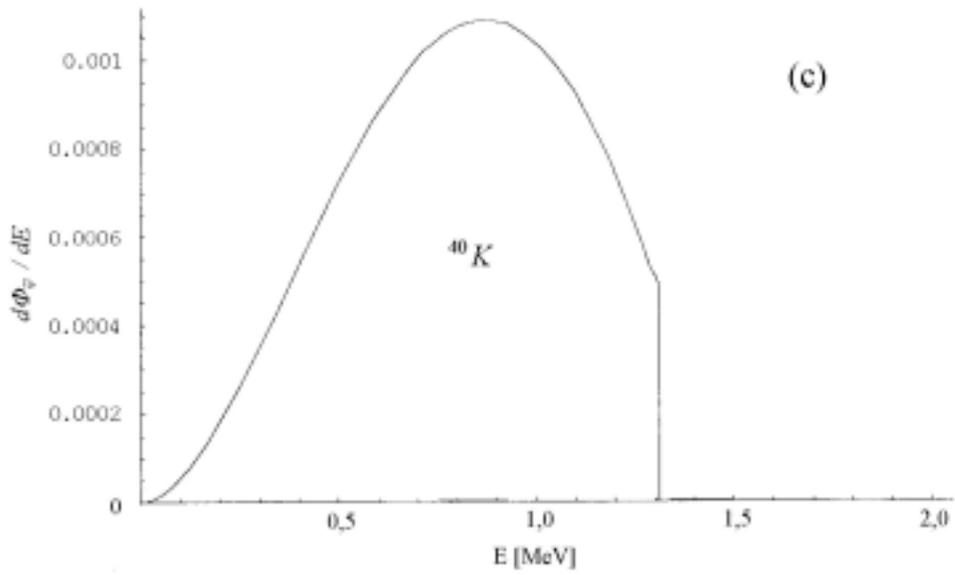

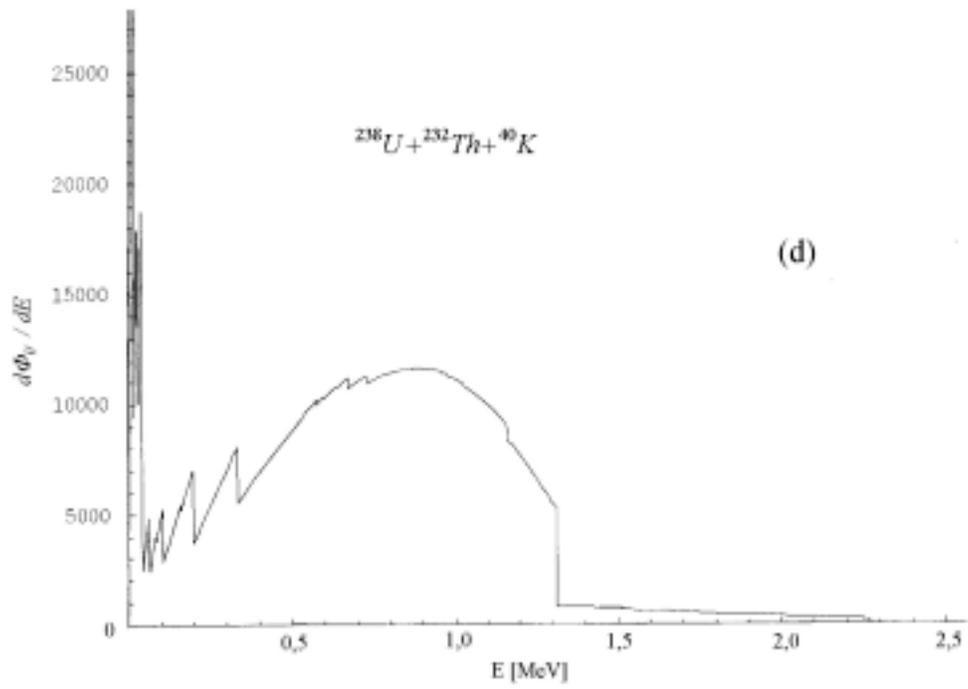

Fig. 3. Calculated partial (a)-(c) and total (d) antineutrino spectra of the Earth. Partial spectra are normalized on nuclear decay and total spectrum normalized on neutrino number per nuclear decay.

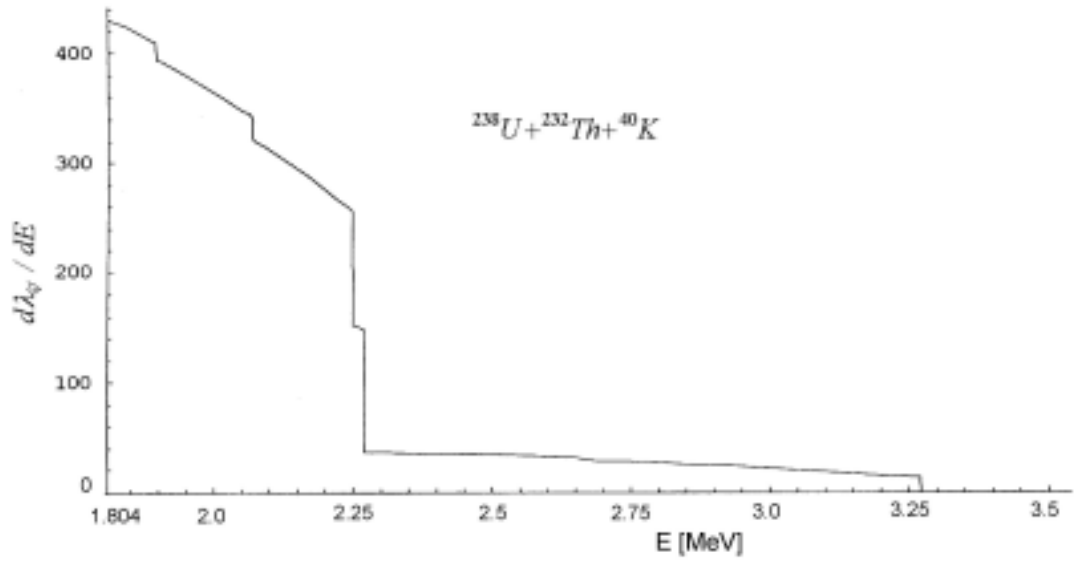

Fig. 4. Total antineutrino spectrum of the Earth with allowance for detection threshold $E_{\tilde{\nu}}^{thr}=1.804$ MeV.

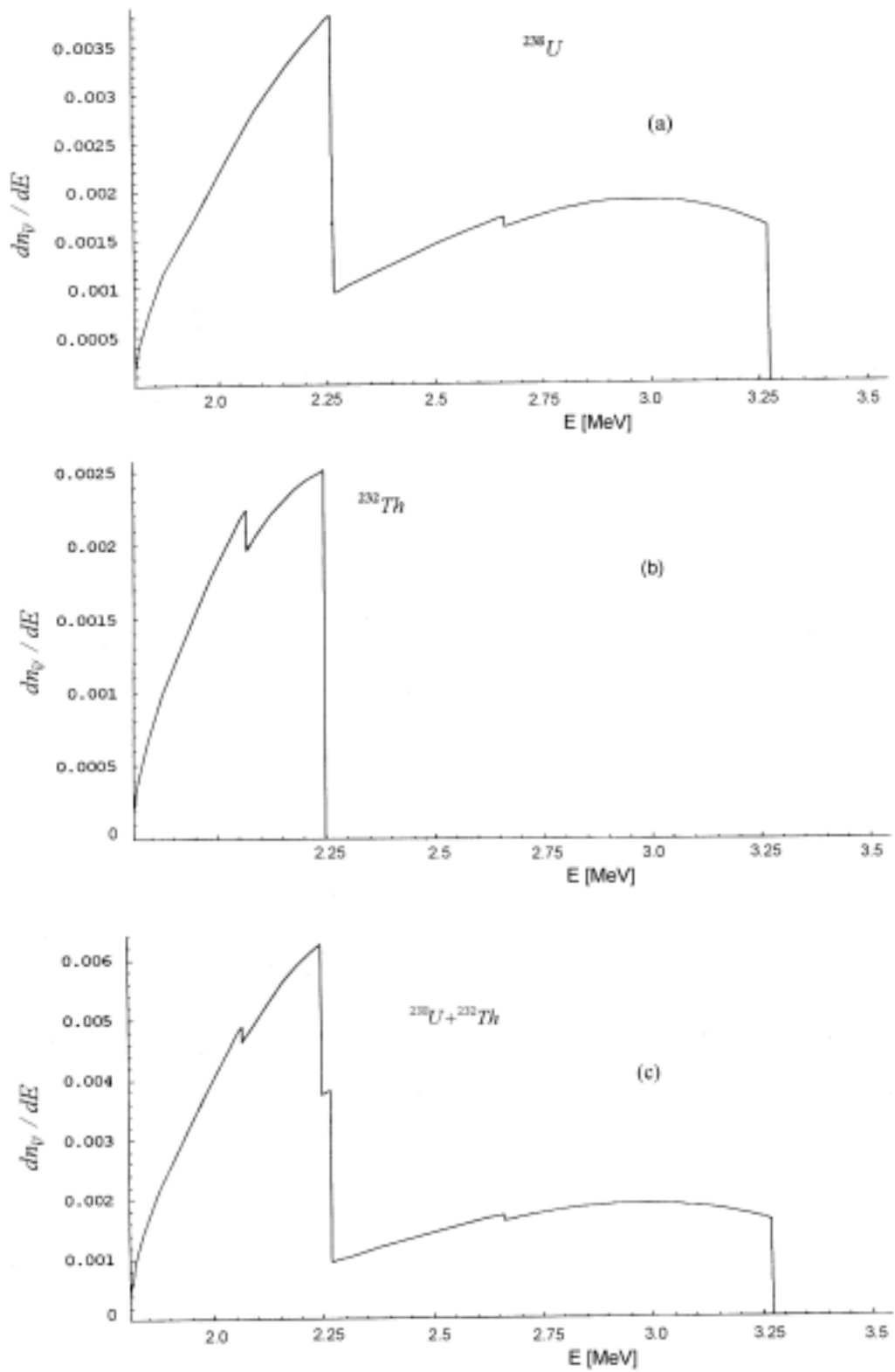

Fig. 5. Calculated partial (a)-(b) and total (c) antineutrino spectra of the Earth in KamLAND detector.